\documentclass{tPHL2e}

\usepackage{epstopdf}
\usepackage{subfigure}

\theoremstyle{plain}

\theoremstyle{definition}

\theoremstyle{remark}

\begin{document}



\title{\textit{Deciphering $M-T$ diagram of shape memory Heusler alloys: Reentrance, plateau and beyond}}

\author{
\name{S. Sergeenkov\textsuperscript{a}$^{\ast}$\thanks{$^\ast$Corresponding author. Email: sergei@df.ufscar.br}, 
C. C\'{o}rdova\textsuperscript{b}, P. Ari-Gur\textsuperscript{c}, V.V. Koledov\textsuperscript{d}, A.P. Kamantsev\textsuperscript{d}, V.G. Shavrov\textsuperscript{d}, A.V. Mashirov\textsuperscript{d}, A.M. Gomes\textsuperscript{e}, A.Y. Takeuchi\textsuperscript{f}, O.F. de Lima\textsuperscript{g}  and  F.M. Ara\'{u}jo-Moreira\textsuperscript{b}}
\affil{\textsuperscript{a}Department of Physics, Universidade Federal da Para\'{\i}ba, Jo\~{a}o Pessoa, PB, Brazil; \\
\textsuperscript{b}Department of Physics, Universidade Federal de S$\tilde{a}$o Carlos, S$\tilde{a}$o Carlos, SP,  Brazil;\\
\textsuperscript{c}Mechanical and Aerospace Engineering, Western Michigan University, Kalamazoo, MI 49008-5343, USA;\\
\textsuperscript{d}Kotelnikov Institute of Radioengineering and Electronics of Russain Academy of Sciences, Moscow 125009, Russia;\\
\textsuperscript{e}Institute of Physics, Universidade Federal do Rio de Janeiro, Rio de Janeiro, RJ, Brazil;\\
\textsuperscript{f}Department of Physics, Universidade Federal do Esp\'{i}rito Santo, Vitoria, ES, Brazil;\\
\textsuperscript{g}Institute of Physics "Gleb Wataghin", UNICAMP, 13083-970 Campinas, SP, Brazi}}

\maketitle

\begin{abstract}
We present our recent results on temperature behavior of magnetization observed in $Ni_{47}Mn_{39}In_{14}$ Heusler alloys. Three regions can be distinguished in the $M-T$ diagram: (I) low temperature martensitic phase (with the Curie temperature $T_{CM}=140K$), (II) intermediate mixed phase (with the critical temperature $T_{MS}=230K$) exhibiting a reentrant like behavior (between $T_{CM}$ and $T_{MS}$), and (III) high temperature austenitic phase (with the Curie temperature $T_{CA}=320K$) exhibiting a rather wide plateau region (between $T_{MS}$ and $T_{CA}$). By arguing that powerful structural transformations, causing drastic modifications of the domain structure in alloys, would also trigger strong fluctuations of the order parameters throughout the entire $M-T$ diagram, we were able to successfully fit all the data by incorporating Gaussian fluctuations (both above and below the above three critical temperatures) into the Ginzburg-Landau scenario. 
\end{abstract}


\section{Introduction}

It is now well established that the origin of many interesting effects in Heusler-based $Ni-Mn-In$ family alloys (including direct and inverse magnetocaloric effects, large magnetic field induced strain, giant magnetoresistance, exchange bias effect, etc.) lies in the intricate coupling between martensitic structural transition and magnetic degrees of freedom of these alloys [1-8]. While high temperature austenitic phase (with the Curie temperature $T_{CA}$) can be readily explained by the interactions between Mn atoms, the origin of low temperature martensitic phase (with the Curie temperature $T_{CM}$), is still rather obscure. One of the possible scenarios for the formation of this phase is based on strong AFM correlations mediated by $N$i atoms [9]. Even more intriguing is the origin of the so-called intermediate phase (lying in-between the above two) exhibiting a marked reentrance-like behavior. As a result of all kinds of magnetic and magnetostructural interactions and correlations, the observable M-T diagram of Heusler-based $Ni-Mn-In$ alloys poses a real challenge for its meaningful interpretation. The main goal of the present paper is to demonstrate that martensitic structural transformation not only causes the drastic modifications of the domain structure and formation of the above-mentioned ordered magnetic phases but that it also triggers strong Gaussian fluctuations of the order parameters (both above and below the critical temperatures) which are ultimately responsible for the diversity of the observed $M-T$ diagram in these alloys. As a typical example of $Ni-Mn-In$ family, we consider here the magnetic response of $Ni_{47}Mn_{39}In_{14}$ alloy.

\section{Samples characterization and magnetic measurements}

\begin{figure}
\centerline{\includegraphics[width=5cm]{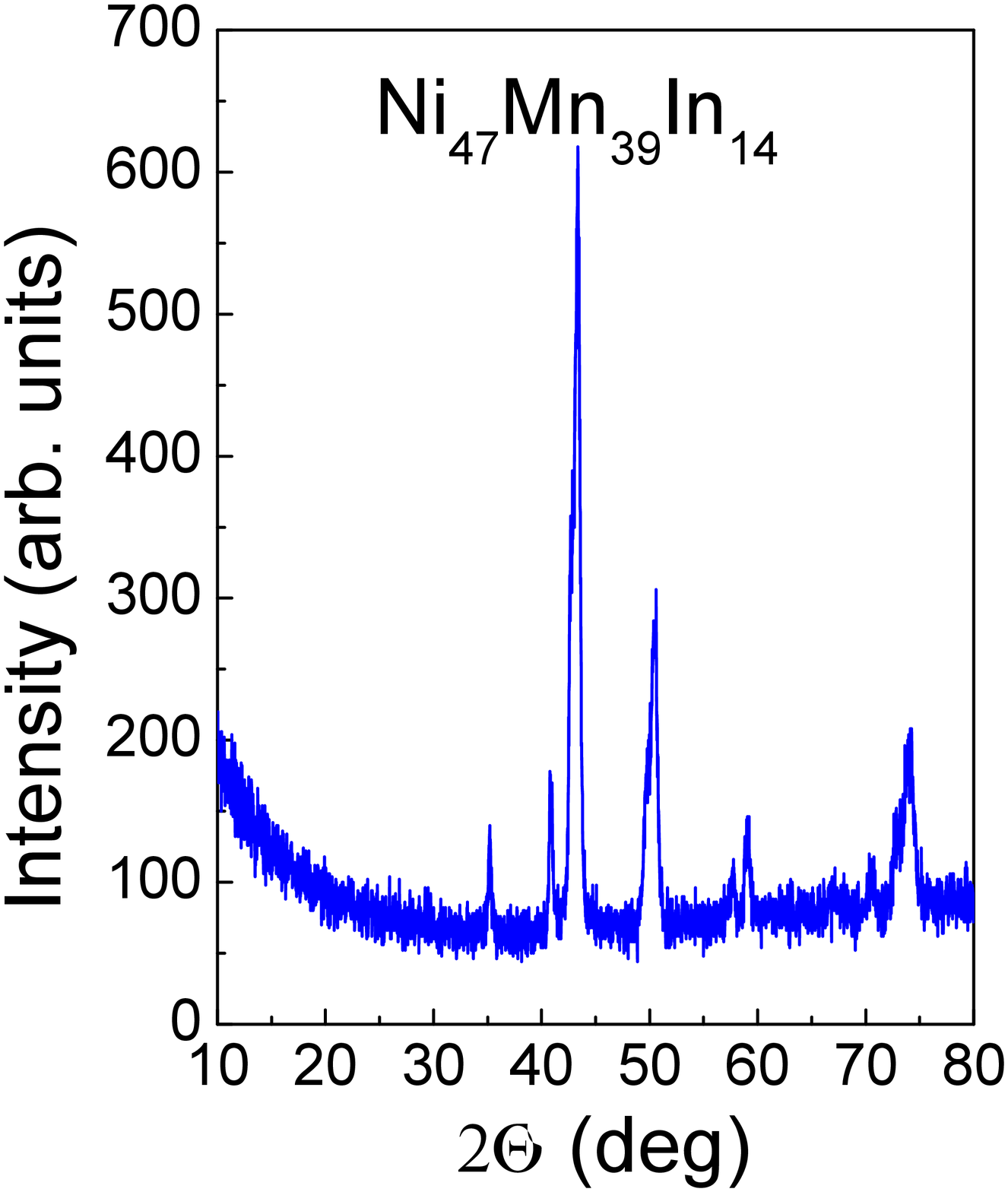}}\vspace{0.1cm}
\caption{XRD pattern taken at room temperature. }
\label{fig:fig1}
\end{figure}

Polycrystalline $Ni_{47}Mn_{39}In_{14}$ samples were prepared from high purity elements ($99.99\%$) by arc melting in argon atmosphere. In order to homogenize the ingots, the samples were annealed in vacuum at $900^{\circ}C$ followed by a gradual cooling to room temperature. In order to study the magnetic thermoelastic properties of the samples, we have cut the plates with the sizes of $3mm \times 12mm \times 1mm$. To probe the structure of the martensitic domains by optical method, the samples were polished with diamond paste. The structural characteristics of our samples were confirmed by X-ray diffraction (XRD). According to XRD pattern (shown in Fig.\ref{fig:fig1}), at room temperature the samples contain cubic (austenite) phase accompanied by low-symmetry phases. Further details about preparation and characterization of our samples can be found elsewhere [10]. 
ZFC and FC measurements of magnetization as a function of temperature $M(T)$ and magnetic field $M(H)$ were performed using a Quantum Design-MPMS-5 SQUID magnetometer working in the temperature range from $2$ to $400K$ and the field range from $0$ to $50kOe$ [10]. 
\begin{figure}
\centerline{\includegraphics[width=5cm]{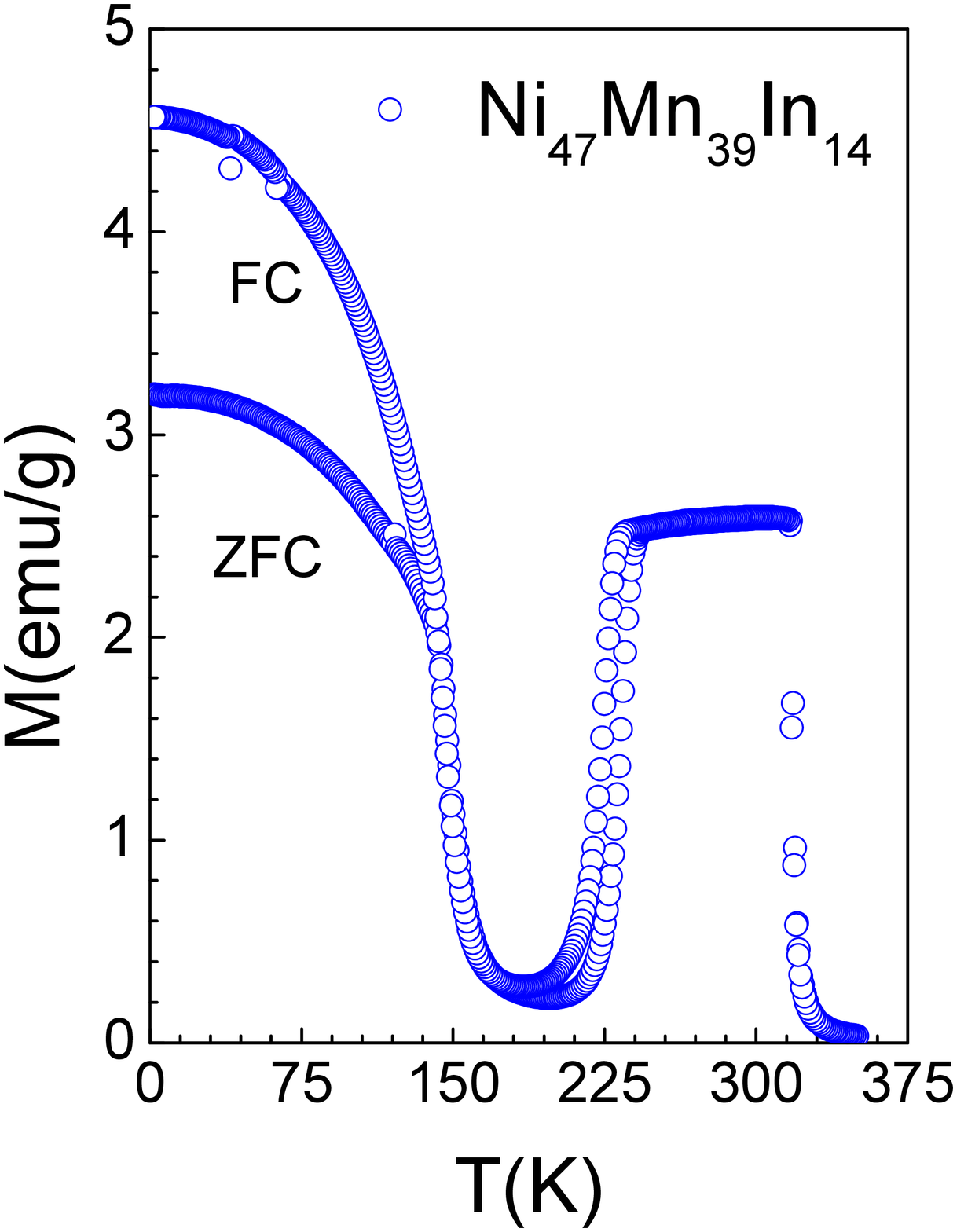}}
\caption{Temperature dependence of ZFC and FC magnetization.  }
\label{fig:fig2}
\end{figure}
\begin{figure}
\centerline{\includegraphics[width=5cm]{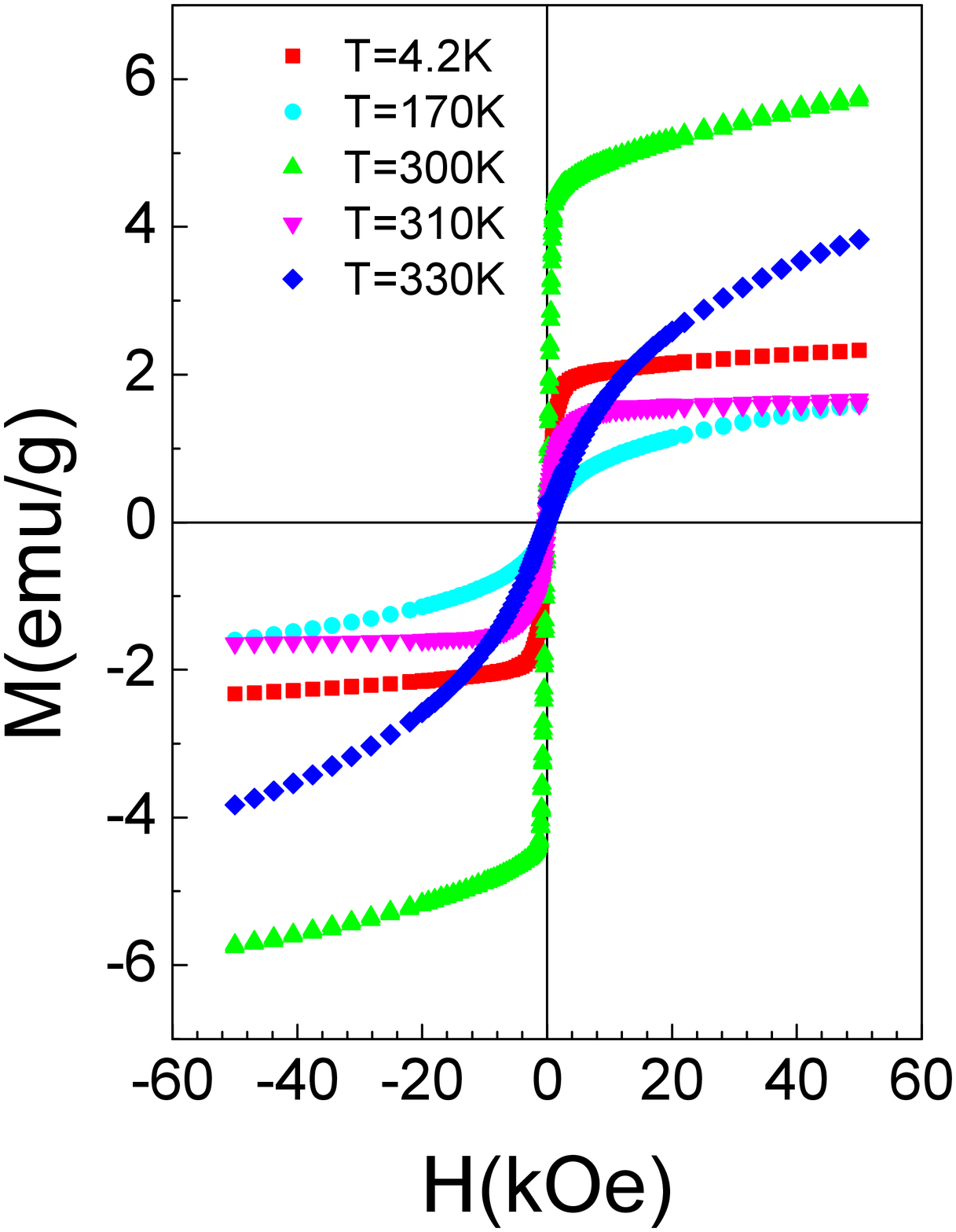}}\vspace{0.1cm}
\caption{$M-H$ curves taken at different relevant temperatures.}
\label{fig:fig3}
\end{figure}

\section{Results and Discussion}

\begin{figure}
\centerline{\includegraphics[width=5cm]{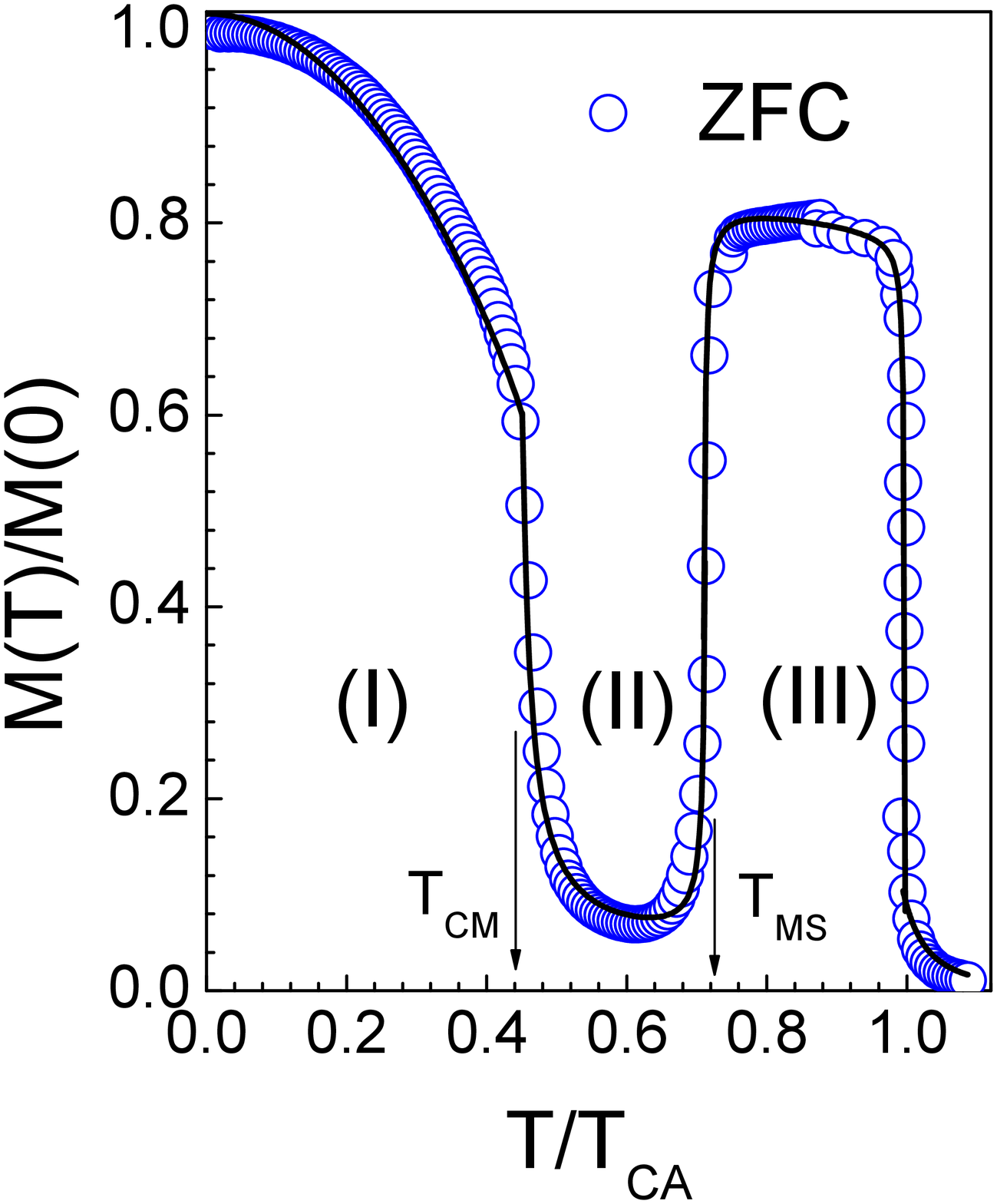}}\vspace{0.1cm}
\caption{The dependence of the normalized ZFC magnetization on normalized temperature. Solid lines are the best fits according to Eqs.(6)-(10).}
\label{fig:fig4}
\end{figure}

Fig.\ref{fig:fig2}  presents the temperature dependence of ZFC and FC magnetization curves for $Ni_{47}Mn_{39}In_{14}$  sample measured in a low magnetic field of $20Oe$. Three regions can be distinguished: (I) low temperature martensitic phase with the Curie temperature $T_{CM}=140K$, (II) reentrant phase (between $T_{CM}=140K$ and $T_{MS}=230K$), and (III) high temperature austenitic phase with the Curie temperature $T_{CA}=320K$. One of the reasons for choosing this particular stoichiometry of the alloy is a rather wide plateau region between $T_{MS}=230K$ and $T_{CA}=320K$. Fig.\ref{fig:fig3}  depicts the measured $M-H$ curves taken at different temperatures relevant to the discussion of the observed $M-T$ diagram. Fig.4 shows the three regions for the normalized ZFC magnetization $M(T)/M(0)$ with $M(0)=3.2emu/g$ as a function of temperature normalized to the austenitic Curie temperature $T_{CA}=320K$. According to Fig.\ref{fig:fig4} , the other two transition temperatures ($T_{CM}$ and $T_{MS}$) are related to $T_{CA}$ as follows: $T_{CM}=0.44T_{CA}$ and $T_{MS}=0.72T_{CA}$.
In order to fully understand the above obtained experimental results, we employ a Ginzburg-Landau (GL) type model for the second order phase transitions. To account for all magnetic changes caused by structural transformations in our sample, we adopt the scenario with multiple order parameters $\eta_i$, by defining the above-introduced three critical temperatures $T_i$ (with $T_1=T_{CM}$, $T_2=T_{MS}$, and $T_3=T_{CA}$) via the vanishing condition $\eta_i(T_i)=0$, separately for three ordered phases marked as $i=1,2,3$, respectively. The resulting GL functional has the following form [11]:
\begin{equation}
F[\eta_i]= a_i\eta_i^2+\frac{1}{2}\beta_i\eta_i^4-\xi_i\eta_i^2  
\end{equation}
Here $a_i(T)=\alpha_i(T^2-T_i^2)$, $\alpha_i$ and $\beta_i$ are the positive constants, and $\xi_i$ are the temperature independent chemical potentials (responsible for residual contributions to magnetization, see below). As usual, the equilibrium state is determined from the minimum-energy condition $\delta F[\eta_i]/\delta \eta_i=0$, which yields for the temperature dependence of the bulk $i$-th order parameter below the corresponding critical temperature $T_i$:
\begin{equation}
\eta_{0i}^2(T)=\frac{\alpha_i(T_i^2-T^2)+\xi_i}{\beta_i}  
\end{equation}                                                   
which, in turn, results in the following expression for the average (mean) value of the magnetization for each of the phases [12]:
\begin{equation}
M_{i,av}(T)\equiv M_{i,s}\eta_{0i}^2(T)=M_{i,0}\left(1-\frac{T^2}{T_i^2}\right)+M_{i,R}
\end{equation}                        
Here, $M_{i,0}\equiv (\alpha_i/\beta_i)T_i^2M_{i,s}$, and $M_{i,R}\equiv (\xi_i/\beta_i)M_{i,s}$ is the temperature independent residual contribution to the average magnetization. $M_{i,s}$ is the corresponding saturation magnetization (which can be deduced from Fig.3). Our preliminary analysis of the experimental data (shown in Fig.4) based on Eq.(3) revealed that, even though it is possible to qualitatively describe the behavior of magnetization in regions (I) and (III), the mean (average) contribution to $M(T)$ utterly fails to explain (even qualitatively) the most interesting reentrance region of the $M-T$ diagram. Therefore, based on our previous experience with manganites (which are shown to exhibit strong Gaussian [11] and critical [12] fluctuations both above and below the Curie temperature), we argue that a similar situation may take place in Heusler alloys as well because rather powerful martensitic transformations drastically modifying the domain structure of the alloys would inevitably trigger strong magnetic fluctuations throughout the entire $M-T$ diagram. That is why, it is quite reasonable to assume that, in order to fully understand the magnetic behavior of our sample, in addition to the average magnetizations of the ordered phases (given by Eq.(3)), fluctuations of their order parameters (leading to fluctuation-induced contributions to magnetization) must be taken into account as well. As we shall demonstrate below (through the final data fits), this rather strong assumption is found to be quite well justified. Recall that according to the  theory of Gaussian fluctuations, the fluctuations of any observable (such as heat capacity, magnetization, etc) which is conjugated to the order parameter $\eta_i$, can be presented in terms of the statistical average of the fluctuation amplitude $<(\delta \eta_i)^2>$ with
$\delta \eta_i =\eta_i -\eta _{0i}$. Namely, the fluctuation-induced contribution to magnetization above ($+$) and below ($-$) the critical point $T_i$ reads [11,13]
\begin{equation}
M_{i,fl}^{\pm}(T)=\frac{M_{i,s}}{Z}\int d\eta_i (\delta \eta_i)^2 e^{-\Sigma [\eta_i]}
\end{equation}
where $Z=\int d\eta_i e^{-\Sigma [\eta_i ]}$ is the partition function with 
$\Sigma [\eta_i ]\equiv ({\cal F} [\eta_i ]-{\cal F} [\eta _{0,i}])/k_BT$. 
Expanding the free energy functional around the mean value of the order parameter $\eta _{0i}$
\begin{equation}
{\cal F} [\eta_i]\approx {\cal F} [\eta _{0,i}]+
\frac{1}{2}\left[ \frac{\partial ^2{\cal F}}
{\partial \eta_i^2}\right ]_{\eta_i =\eta _{0,i}}(\delta \eta_i )^2
\end{equation}                              
we can explicitly calculate the Gaussian integrals. Due to the fact that $\eta _{0i}$ is given by Eq.(2) below $T_i$ and vanishes at $T>T_i$, we obtain finally
\begin{equation}
M_{i,fl}^{-}(T)=\left(\frac{1}{3}\right)\frac{M_{i,R}^2}{M_{i,0}\left(1-\frac{T^2}{T_i^2}\right)-M_{i,R}},
\qquad T\le T_i
\end{equation}
and
\begin{equation}
M_{i,fl}^{+}(T)=\left(\frac{2}{3}\right)\frac{M_{i,R}^2}{M_{i,0}\left(\frac{T^2}{T_i^2}-1\right)+M_{i,R}},
\qquad T\ge T_i
\end{equation}
for the fluctuation-induced contributions to magnetization below and above the critical temperature $T_i$, respectively. It is instructive to point out that, according to Eqs.(6) and (7), the temperature dependent fluctuations are actually driven by residual contributions $M_{i,R}$. Notice also that the total magnetization below $T_i$ is given by a combination of the average and fluctuation contributions, $M_{i}^{-}=M_{i,av}+M_{i,fl}^{-}$, while above $T_i$, it is completely determined by the fluctuations, that is $M_{i}^{+}=M_{i,fl}^{+}$. Besides, in view of Eqs.(3)-(7), at the critical temperature (when $T=T_i$), we have $M_{i}^{-}(T_i)=M_{i,R}+M_{i,fl}^{-}(T_i)=M_{i,R}-\frac{1}{3}M_{i,R}=\frac{2}{3}M_{i,R}$ which indeed coincides with $M_{i}^{+}(T_i)=M_{i,fl}^{+}(T_i)=\frac{2}{3}M_{i,R}$ and hence provides the necessary self-consistency of the adopted in this paper approach. Based on the previous discussion, we can describe now the temperature dependence of the observed magnetization for each of the three regions (the best fits are shown by solid lines in Fig.4 with $M_{1,0}=1.7emu/g$, $M_{2,0}=1.6emu/g$, $M_{3,0}=2.4emu/g$, $M_{1,R}=0.8emu/g$, $M_{2,R}=0.9emu/g$, and $M_{3,R}=0.1emu/g$):
\begin{eqnarray}
 &&  M_I(T)=M_{1,av}(T)+M_{1,fl}^{-}(T)\\ 
 && M_{II}(T)=M_{2,av}(T)+M_{2,fl}^{-}(T)+M_{1,fl}^{+}(T)\\ 
 && M_{III}(T)=M_{3,av}(T)+M_{3,fl}^{-}(T)+M_{2,fl}^{+}(T) 
\end{eqnarray}          
And finally, as expected, a short tail seen above the austenitic phase transition temperature $T_{CA}$ is readily fitted by the fluctuation induced term $M_{3,fl}^{+}$ driven by a small residual contribution $M_{3,R}$. 
We would like to draw a special attention to rather large values of the residual contributions $M_{1,R}$ and $M_{2,R}$ (in comparison with $M_{3,R}$) because, to a large degree, they are responsible for the two marked features in the observed $M-T$ diagram. First of all, according to Eq.(9), they actively participate in the appearance of reentrant like phase which is fuelled by strong fluctuations induced by $M_{1,R}$ and $M_{2,R}$ terms above $T_{CA}$ and below $T_{MS}$, respectively. And secondly, in view of Eq.(10), strong fluctuations above $T_{MS}$ (driven by the residual contribution $M_{2,R}$) play a crucial role in the formation of the plateau region (in-between $T_{MS}$ and $T_{CA}$, see Fig.4). It can be directly verified that the width of this plateau indeed depends on the absolute value of $M_{2,R}$. Comparing the magnetic response of this sample with the performance of other alloys prepared with different concentration of $In$, we are able to conclude that the value of the residual contribution $M_{2,R}$ decreases with the decrease of $In$ content in our samples. More precisely, the samples with $In$ content below $12\%$ (and hence with smaller values of $M_{2,R}$) exhibit a much more narrow plateau and, at the same time, a much larger reentrant region which may extend from $T_{MS}$ all the way up to $T_{CA}$ [1,6,7,9].

\section{Conclusions}

By analyzing our experimental data on the temperature dependence of magnetization in $Ni_{47}Mn_{39}In_{14}$ Heusler alloys, we came to the conclusion that they can be quite well understood assuming the presence (in addition to ordered magnetic phases) of strong magnetic fluctuations of the order parameters triggered by powerful structural transformations. Accounting for these Gaussian fluctuations (within the GL scenario) allowed us to explain the origin of the observed reentrance and plateau regions of the $M-T$ diagram. 

\section*{Acknowledgements}

We are indebted to Marcel Ausloos (Liege and Leicester) and Alex Kuklin (Dubna) for very useful discussions. The authors gratefully acknowledge Brazilian agencies FAPESQ (DCR-PB), FAPESP, CNPq and CAPES for financial support. In addition, this study was supported by Award No. RUP1-7028-MO-11 of the US Civilian Research \& Development Foundation (CRDF), RSF (Grant No 14-22-00279) and by the National Science Foundation under Cooperative Agreement No. OISE-9531011.


\begin{thebibliography}{99}

\bibitem{1} Thorsten Krenke, Mehmet Acet, Eberhard F. Wassermann, Xavier Moya, Lluis Manosa,  and Antoni Planes, Phys. Rev. B \textbf{73} (2006) p. 174413.

\bibitem{2} 	V.K. Sharma, M.K. Chattopadhyay, Ravi Kumar, Tapas Ganguli, Pragya Tiwari,  and S.B. Roy, J. Phys.: Condens. Matter \textbf{19} (2007) p. 496207.

\bibitem{3}	Zhe Li, Chao Jing, Jiping Chen, Shujuan Yuan, Shixun Cao, and Jincang Zhang, Appl. Phys. Lett. \textbf{91} (2007) p. 112505.
 
\bibitem{4}  S. Chatterjee, S. Giri, S. Majumdar, and S.K. De, J. Phys. D \textbf{42} (2009) p. 065001.

\bibitem{5}	S. Chatterjee, S. Giri, S.K. De, and S. Majumdar, Phys. Rev. B \textbf{79} (2009) p. 092410.
 
\bibitem{6}	Lluis Manosa, David Gonzalez-Alonso, Antoni Planes, Erell Bonnot, Maria Barrio, Josep-Lluis Tamarit, Seda Aksoy, and Mehmet Acet, Nature Materials \textbf{9} (2010) p. 478.

\bibitem{7}	A.P. Kazakov, V.N. Prudnikov, A.B. Granovsky, A.P. Zhukov, J. Gonzalez, I. Dubenko, A.K. Pathak, S. Stadler,  and N. Ali, Appl. Phys. Lett. \textbf{98} (2011) p. 131911.

\bibitem{8}	V. Vega, L. Gonzalez, J. Garcıa, W.O. Rosa, D. Serantes, V.M. Prida, G. Badini, R. Varga, J.J. Sunol,  and B. Hernando, J. Appl. Phys. \textbf{112} (2012) p. 033905.

\bibitem{9}	K.R. Priolkar, P.A. Bhobe, D.N. Lobo, S.W.D. Souza, S.R. Barman, Aparna Chakrabarti, and S. Emura, Phys. Rev. B \textbf{87} (2013) p. 144412.

\bibitem{10} E.C. Passamani, C.  C\'{o}rdova, A.L. Alves, P.S. Moscon, C. Larica, A.Y. Takeuchi, and A. Biondo, J. Phys. D \textbf{42} (2009) p. 215006.

\bibitem{11} S. Sergeenkov, H. Bougrine, M. Ausloos, and A. Gilabert, Phys. Rev. B \textbf{60} (1999) p. 12322.

\bibitem{12} S. Sergeenkov, H. Bougrine, M. Ausloos and A. Gilabert, JETP Letters \textbf{70} (1999) p. 141.

\bibitem{13}	 S. Sergeenkov, C.A. Cardoso, M.R.B. Andreeta, A.C. Hernandes, E.R. Leite, and F.M. Araujo-Moreira, Physica Status Solidi A \textbf{208} (2011) p. 1704. 
 

\end{thebibliography}
\end{document}